\documentclass[final,5p,times,twocolumn]{elsarticle}

\usepackage{amssymb}

\journal{Geological Society of America}

\begin{document}

\begin{frontmatter}


 \author{Nicola Scafetta \corref{cor1}\fnref{label2}}
 \ead{ns2002@duke.edu}

\title{Total solar irradiance satellite composites and
their phenomenological effect on climate}


\address{Department of Physics, Duke University, Durham, NC 27708, USA.}

\begin{abstract}
\textbf{Herein I discuss and  propose updated satellite composites  of the total  solar irradiance covering the
period 1978-2008. The composites are compiled from measurements made with the three ACRIM experiments. Measurements from the
NIMBUS7/ERB and the ERBS/ERBE satellite experiments are used to fill the gap from June 1989 to
October 1991 between ACRIM1 and ACRIM2 experiments. The climate implications of the  alternative satellite composites are discussed by using a phenomenological climate model for reconstructing the total solar irradiance signature on climate during the last four centuries.}
\end{abstract}

\begin{keyword}
This paper has been presented at the 2007 GSA Denver Annual Meeting(28–31 October 2007). Session n. 187: \emph{The Cause of Global Warming. Are We Facing Global Catastrophe in the Coming Century?}, Colorado Convention Center: 605/607 8:00 AM-12:00 PM, Wednesday, 31 October 2007. This paper is currently in press on a special GSA volume dedicated to the conference session. \emph{http://gsa.confex.com/gsa/2007AM/finalprogram/session\_19366.htm}

This paper has substituted the scheduled presentation by Richard Willson ``Variations of total solar irradiance and their implication for climate change'' in Geological Society of America Abstracts with Programs, Vol. 39, No. 6, p. 507.
\emph{http://gsa.confex.com/gsa/2007AM/finalprogram/abstract\_130944.htm}




\end{keyword}

\end{frontmatter}



\section{Introduction}

A contiguous TSI database of satellite  observations extends from
late 1978 to the present, covering  30 years, that is, almost three
sunspot 11-year cycles. This database is  comprised of the
observations of seven independent experiments: NIMBUS7/ERB [Hoyt \emph{et al.}, 1992],
SMM/ACRIM1 [Willson  and  Hudson, 1991], ERBS/ERBE [Lee III \emph{et al.}, 1995], UARS/ACRIM2 [Willson, 1994; Willson, 1997], SOHO/VIRGO [Fr\"ohlich \emph{et al.}, 1997; Crommelynck  and  Dewitte, 1997], ACRIMSAT/ACRIM3 [Willson, 2001]. There exists another TSI satellite record,
SORCE/TIM [Kopp \emph{et al.}, 2003], but it is not studied here because it started just on February 2003, and it is still too short for our purpose. None of these independent data sets  cover the entire
period of observation, thus a composite of the database is necessary
to obtain a consistent picture about the TSI variation. Herein, we
use the records plotted in Figure  1.

Three TSI satellite composite are currently available:  the ACRIM composite
[Willson and Mordvinov, 2003], the PMOD composite [Fr\"ohlich and Lean, 1998;
 Fr\"ohlich, 2000, 2006] and the IRMB composite [Dewitte \emph{et al.}, 2004],
 respectively: see Figure 2. Each composite is compiled by using
different models corresponding to different mathematical
philosophies and different combinations of data.

For example, one of the most prominent differences between ACRIM and
PMOD composites  is due to the different way of how the two teams
use the NIMBUS7/ERB record to fill the period 1989.53-1991.75, the
so-called ACRIM-gap between ACRIM1 and ACRIM2 records. The
consequence is that these two composites significantly differ from
each other, in particular about whether the minimum of the TSI during
solar cycle 22-23 (1995/6) is approximately $0.45 W/m^2$ higher
(ACRIM composite) or approximately at the same level (PMOD) as the
TSI minimum during solar cycle 21-22 (1985/6). Figure 3 shows the difference between ACRIM and PMOD.

\begin{figure}\label{AAAA}
\includegraphics[angle=-90,width=20pc]{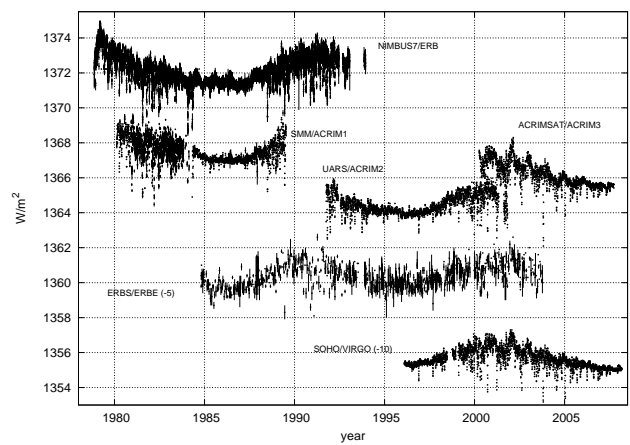}
 \caption{ TSI satellite records.  ERBS/ERBE and SOHO/VIRGO TSI records are shifted by $-5W/m^2$ and $-10W/m^2$, respectively, from the `native scale' for visual convenience. (units of watts/meter$^2$ at 1 A.U.)
 }
\end{figure}

The difference among the TSI satellite composites has significant
implications not only  on solar physics where  the correctness of
the theoretical models must be necessarily tested against the actual
observations, and not vice versa, but also on the more general global
warming debate. Phenomenological analyses [Scafetta and West,
2007,2008]  using TSI proxy, satellite composites and  global surface temperature records
of the past 400 years  show that solar
variation has been a dominant forcing for climate change during both
the pre- and industrial era. According to these analyses, the
sun will likely be a dominant
contributor to climate change in the future.
 However, the solar contribution to the global
warming during the last three decades remains severely uncertain due
mostly to the difference between the TSI satellite composites.

\begin{figure}\label{AIP}
\includegraphics[angle=-90,width=20pc]{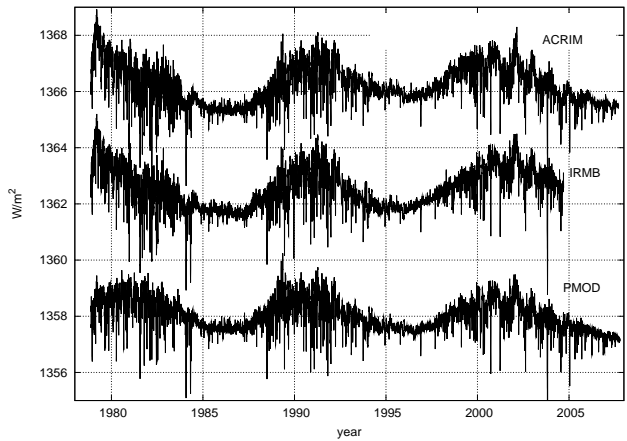}
 \caption{
 ACRIM, IRMB and PMOD TSI satellite composites.
 }
\end{figure}

 The phenomenological solar signature  on the global temperature is found to match quite
well 400 years of data since 1600 [Scafetta and West, 2007-2008],
but such an almost continuous matching would be abruptly interrupted since
1975 if the PMOD composite is adopted. Instead, by adopting the
ACRIM composite it is still possible to notice  a significant correlation
between temperature data and the reconstruction of the solar effect
on climate  (see figure 6 in Scafetta and West [2007] in and figure
1 in Scafetta and West [2008]). Thus, a significant fraction of the
+0.4$K$ warming observed from 1980 to 2007 can be ascribed to an
increase of the solar activity if ACRIM composite is adopted, but
almost none of it would be linked to solar activity if PMOD
composite is adopted. Evidently, if the solar contribution is
uncertain, the anthropogenic contribution to the global warming during
the last three decades  is uncertain as well. Hence, determining the
correct TSI composite during  the last three decades should be
considered of crucial importance.

Note that the climate models adopted by the Intergovernmental Panel
on Climate Change [IPCC, 2007] do not agree with the above
phenomenological findings and predict but a minor solar contribution
to climate change during the last century and, in particular, during the last 30 years. However,  such
climate models assume that  the TSI forcing is the only solar forcing of climate and use as TSI record the one derived from the TSI proxy
reconstructions proposed by Lean [Lean, 2000; Wang, 2005] which are
compatible with the PMOD TSI composite since 1978. This becomes problematic if PMOD TSI composite is found to be flawed. In any case, the small climate sensitivity to solar  changes predicted by the current climate models is also believed to be due to the absence of several climate feedback mechanisms that may be quite sensitive to solar changes, in  addition to TSI changes alone. Some of these phenomena include, for example, the UV modulation of ozone concentration that would effect  the stratosphere water vapor feedback and the modulation of the cloud cover due to the variation of cosmic ray flux which is linked to  changes of the magnetic solar activity [Pap \emph{et al.}, 2004; Kirkby , 2007]. These climate mechanisms  are expected to magnify the influence of a solar change  on climate.

The original ACRIM composite [Willson and Mordvinov, 2003] has been
constructed by simply calibrating the three ACRIM datasets and the
NIMBUS7/ERB record on the base of direct comparison of the
\emph{entire} overlapping region between two contiguous satellite
records. This composite does not alter the actual observations as
they have  been published by the original experimental groups.
However, if some  degradation or glitches do exist in the data, this
composite is flawed for at least two reasons: 1) the mathematical
methodology used for merging the two contiguous satellite records, which
uses just the average during the \emph{entire} overlapping regions between  two records, may
easily give biased estimates;  2) if the NIMBUS7/ERB record
presents some glitches, or degradation did occur during the ACRIM-gap,
the relative position of ACRIM1 and ACRIM2 is falsified.

The IRMB composite [Dewitte \emph{et al.}, 2004] is constructed by
first referring all datasets to space absolute radiometric
references, and then the actual value for each day is obtained by
averaging all available satellite observations for that day. Thus,
IRMB composite  adopts a statistical
average approach among all available observations; evidently, because the daily
average estimate is based on a small set of data (1, 2 or in a few cases
3 data per day), it is not statistically robust, and this may easily
produce artificial slips every time data from a specific record are missing
or added.

The PMOD composite [Fr\"ohlich and Lean, 1998; Fr\"ohlich, 2000,
2004, 2006]  is constructed by altering the published experimental
TSI satellite data every time the PMOD team claims that the published data are
corrupted because  of presumed sudden glitches due to changes in
the orientation of the spacecraft and/or to switch-offs of the
sensors, or because of some kind of instrumental degradation. Some
TSI theoretical model predictions [Lee III \emph{et al.}, 1995;
Chapman \emph{et al.}, 1996; Fro\"ohlich and Lean, 1998] have been
heavily used by the PMOD team to identify, correct and evaluate
these presumed errors in the published TSI satellite records, and these models have been
changed constantly during the last 10 years.

PMOD composite is
claimed to be  consistent with some  TSI theoretical proxy models
 [Wenzler \emph{et al.}, 2006;
Krivova \emph{et al.}, 2007]. However, differences between the model
and the PMOD TSI composite can be easily recognized: for example,
Wenzler \emph{et al.} [2006] need to calibrate the model on the PMOD
composite itself to improve the matching, and several details are not
reproduced. Also it can not be excluded that an alternative calibration of the parameters of these TSI proxy models may better fit the ACRIM TSI satellite composite. Evidently, if the above theoretical models and/or the
corrections of the satellite records implemented by the PMOD team
are found to be severely flawed, PMOD is flawed as well. In any case,
an apparent agreement between some theoretical TSI model which depends on several calibration parameters and a TSI satellite composite does
not necessarily indicate the correctness of the latter because in
science theoretical models should be tested and evaluated against
the actual observations, and not vice versa.

Herein, we  construct  alternative TSI satellite composites using an approach similar to that adopted by the ACRIM team, that is, we  do not alter the published satellite data by using predetermined theoretical models that may bias the composite. However, contrary to the original ACRIM team's approach we use a methodology that takes into account the evident statistical biases that are found in the published satellite records. The three ACRIM records  are preferred and the ACRIM gap is filled by using the measurements from the
NIMBUS7/ERB and the ERBS/ERBE satellite experiments.  Finally, we use these alternative TSI satellite composites in conjunction with a recent TSI proxy reconstruction proposed by  Solanki's team [Krivova \emph{et al.}, 2007] to reconstruct the signature of solar change on global climate  using a phenomenological model [Scafetta and West, 2007, 2008].

\begin{figure}\label{A-P}
\includegraphics[angle=-90,width=20pc]{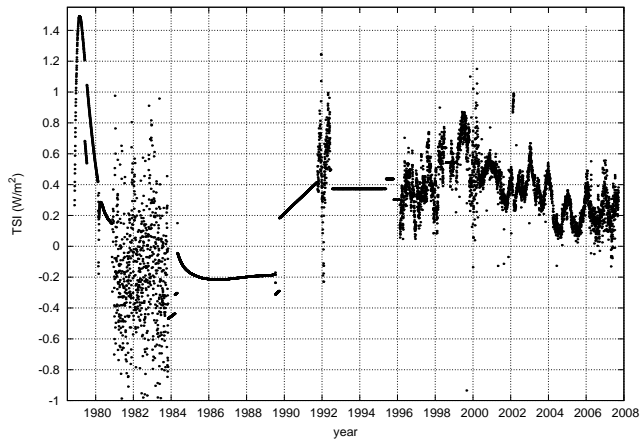}
 \caption{
Relative difference between  ACRIM and PMOD TSI satellite composites.
 }
\end{figure}

\section{SMM/ACRIM1  vs.  NIMBUS7/ERB}

The first step is to compose the SMM/ACRIM1  and the NIMBUS7/ERB
records. Note that the
relative accuracy, precision and traceability of these two databases
are radically different. In particular, the average error of
NIMBUS7/ERB measurements is $\pm 0.16 W/m^2$ while the average error
of SMM/ACRIM1  measurements is $\pm 0.04 W/m^2$: thus SMM/ACRIM1 is
significantly more precise than NIMBUS7/ERB. Moreover, NIMBUS7/ERB
was not able to continuously calibrate its sensor degradations as ACRIM1 was. NIMBUS7/ERB radiometer was calibrated electrically every 12 days.
For the above reasons  ACRIM1 measurements are supposed to be more
accurate than the NIMBUS7/ERB one and, when available, they are
always preferred to the  NIMBUS7/ERB ones.

It is necessary to adopt the NIMBUS7/ERB record for
reconstructing the TSI record during three periods: before
17/02/1980,  from 04/11/1983 to  03/05/1984,  and
after 14/07/1989. To accomplish this we  evaluate the position of
NIMBUS7/ERB relative to SMM/ACRIM1: we plot this in Figure
4. The black smooth curve is a  91-day moving average.

\begin{figure}\label{N7-A1}
\includegraphics[angle=-90,width=20pc]{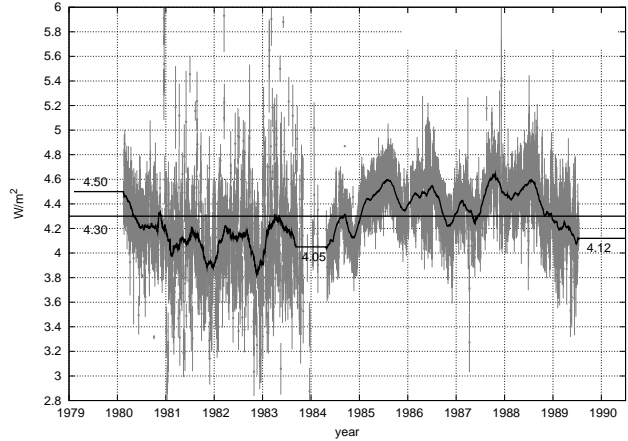}
 \caption{
 Relative difference between NIMBUS7/ERB and SMM/ACRIM1 TSI records.
 }
\end{figure}

The data shown in Figure 4 have an average of $4.3 W/m^2$.
The original ACRIM TSI  composite [Willson and Mordvinov, 2003] is constructed in a way which is equivalent to
use the above average value to merge SMM/ACRIM1  and  NIMBUS7/ERB
record.

However,  if the differences between SMM/ACRIM1  and the
NIMBUS7/ERB  records were only due to random fluctuations around an
average value,  the error associated to the 91-day moving average
values had to be about $\pm 0.02$-$0.03 W/m^2$, as calculated from
the measurement uncertainties. Because the standard deviation of the smooth
data shown in Figure 4 is significantly larger, $\pm 0.17
W/m^2$, the difference between SMM/ACRIM1  and  NIMBUS7/ERB
measurements is not just due to random fluctuations, but  due
to biases and trends in the data probably due to poor sensor calibration of NIMBUS7/ERB.

Under the theoretical assumption that  the SMM/ACRIM1 measurements are more accurate than NIMBUS7/ERB,  the
black  91-days moving average smooth curve shown in Figure 4
suggests that NIMBUS7/ERB  measurements could gradually shift during a relatively short period of time, a few months, by
an amount that on average is about $0.2 W/m^2$, and   in a few
cases can also be as large as $0.5 W/m^2$. In fact, irregular large
oscillations with periods ranging from 5 to 12 months are clearly
visible in Figure 4.

The above finding shows  that the original methodology adopted by the ACRIM
team to merge SMM/ACRIM1  and  NIMBUS7/ERB records is likely inappropriate
because it assumes that NIMBUS7/ERB data are statistically
stationary, that is, unbiased, while this is not what is found in
the data. By not taking into account this problem  the ACRIM team's
methodology can introduce significant artificial slips in the TSI
satellite composite at the chosen merging day.

To reduce the errors due to the above irregular large oscillations
of NIMBUS7/ERB  measurements
 we used the  black 91-days moving average smooth
 curve shown in  Figure 4 to  reduce NIMBUS7/ERB record
 to the level of SMM/ACRIM1 during the  overlapping period and,
 thus, use the NIMBUS7/ERB corrected record to fill all days SMM/ACRIM1
 record misses. Before 17/02/1980 NIMBUS7/ERB record is shifted
  by  $-4.5 W/m^2$, while  after 14/07/1989 it is shifted by $-4.12 W/m^2$. This means that relative to the original ACRIM composite, our composite will be $0.2 W/m^2$ lower before 17/02/1980, and $0.18 W/m^2$ higher  after 14/07/1989.

However, these values do depend on the  moving average window adopted. In fact, by increasing the window the two above
levels will  approach to the average level at $-4.3
W/m^2$, which is the value  used by the ACRIM team to merge the two records. Thus, the above two estimates can have an
error as large as $0.2 W/m^2$.

On the contrary, the PMOD team significantly alters both SMM/ACRIM1  and
NIMBUS7/ERB records before 1986: see Figure 3. These corrections are not justified by the data themselves, but by
theoretical models, which can be erroneous and/or may have large uncertainties.
About the  SMM/ACRIM1
data PMOD team  assumes that the  SMM/ACRIM1 record from 1984 to 1986
significantly degradated: as Figure 3 shows, the
position of   NIMBUS7/ERB relative to SMM/ACRIM1 gradually increases
from  1984 to 1986. However, this pattern can be caused both by a degradation of
SMM/ACRIM1, as the PMOD team interprets, and by an increase of sensitivity of NIMBUS7/ERB sensors due to undetermined factors. In fact, we
observe that from 1988 to 1989.5  the  position of   NIMBUS7/ERB
relative to SMM/ACRIM1 gradually decreased of  the same amount
of the gradual increase observed from  1984 to 1986. This suggests that
NIMBUS7/ERB record of the TSI can gradually vary by these
large amounts.

In any case,  ACRIM team has never published an update of their
SMM/ACRIM1 record and publicly disagreed with the PMOD team on many occasions on this issue
[Willson and Mordvinov, 2003] because they were not able to find any physical explanation for this presumed degradation of the SMM/ACRIM1 record. Herein,
we believe that the ACRIM team opinion cannot be just ignored and dismissed given
the fact they are the authors of the data. The correction
implemented by the PMOD team on the SMM/ACRIM1 record should be
considered hypothetical and not taken as granted. In any case, this
correction would not alter the position of the TSI minimum in
1985/1986 relative to the minimum in 1996, which herein is a more
important issue; PMOD team's correction would only lower  the TSI
maximum in 1981/1982 by about $0.2W/m^2$    relative
to the TSI satellite composite after 1986.

About  the NIMBUS7/ERB record before 1980,  although
NIMBUS7/ERB trends appear to be quite uncertain, the PMOD team's correction of them
should be considered hypothetical as well because they are not
justified by other satellite measurements. In particular, PMOD team
believes that the large NIMBUS7/ERB peak occurred during the first
months of 1979 (see Figures   1  and 2) is an
artifact due to changes in the orientation of the spacecraft that
has to be corrected. However, we observe that TSI theoretical
reconstruction proposed by Solanki [Wenzler \emph{et al.}, 2006]
shows that a large TSI peak  occurred during the first months of
1979. Look carefully at their figures 14 and 15 where the TSI proxy
reconstruction is compared with the PMOD composite; during the first
months of 1979 there is a discrepancy of about $1 W/m^2$ between the
two records. This is a sufficient evidence for considering the PMOD
team's corrections of NIMBUS7/ERB suspicious.

In any case, the exact TSI patterns before 17/02/1980 and during
the ACRIM-gap should be considered highly uncertain because they have to be derived from low quality satellite measurements.

\section{The ACRIM-gap: 15/07/1989 - 03/10/1991}

 SMM/ACRIM1 and UARS/ACRIM2 records can only be bridged by using two
 low quality satellite records: NIMBUS7/ERB and ERBS/ERBE.
Figure 5 shows the two records. Note that  NIMBUS7/ERB
and ERBS/ERBE present opposite trend. From 1990 to
1991.5   NIMBUS7/ERB record  shows an increasing trend while
ERBS/ERBE record shows a decreasing trend [Willson and Mordvinov,
2003]. Thus, the only two available satellite records are not
compatible with each other and at least one of the two is
corrupted. Note that ERBS/ERBE too was unable
to calibrate its sensor degradations and a direct comparison with the ACRIM records reveals that
the discrepancy between local ACRIM smooth trends and the ERBS/ERBE smoth trends may as large as $\pm 0.2 W/m^2$: the amplitude of these non-stationary biases is smaller than that observed in the NIMBUS7/ERB  measurements, but they are still significant. Moreover, the average error of
ERBS/ERBE's measurements are the largest  among all satellite
observations: $\pm 0.26 W/m^2$.

\begin{figure}\label{N7-ERBS}
\includegraphics[angle=-90,width=20pc]{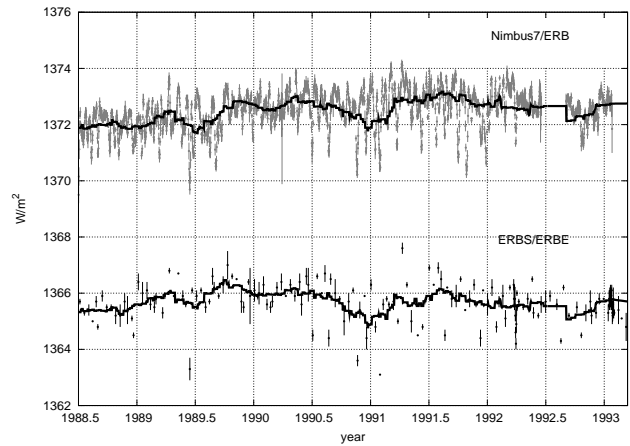}
 \caption{
 NIMBUS7/ERB and ERBS/ERBE TSI records during the ACRIM-gap. The smooth curves are 91-day moving averages calculated on the TSI values only for those days available in both records.
 }
\end{figure}

The PMOD team claims that NIMBUS7/ERB record must be severely
corrected during the ACRIM-gap.
The reasons of these corrections can be found in the literature. Lee
III  \emph{et al.} [1995] compared the NIMBUS7/ERB dataset with a
TSI proxy model based on a multi-regression analysis of March 1985 to
August 1989 ERBS/ERBE irradiance measurements. They concluded that
after September 1989 NIMBUS7/ERB time series appeared to abruptly
increase by $+0.4 W/m^2$ after a switch-off of NIMBUS7/ERB  for four
days. Another $+0.4 W/m^2$ upward shift  appeared to  occur on
April 1990. Thus, this model suggests a two step shift correction that, once combined,
would require the NIMBUS7/ERB record to be shifted down by $0.8 W/m^2$ at the end
of April 1990.

Later Chapman et al. [1996] review the finding by Lee \emph{et al.} [1995]
and concluded that on 29/09/1989 there was an upward shift of $+0.31
W/m^2$ and on 9/05/1990 there was an upward shift of $+0.37 W/m^2$:
the new proposed combined two step correction shift had to be -0.68 $W/m^2$ after 9/05/1990. Afterward,  Fr\"ohlich and Lean
[1998] suggested a different two step shift correction, that is, the NIMBUS7/ERB record had to be adjusted by $-0.26W/m^2$
and $-0.32 W/m^2$ near October 1, 1989, and May 8, 1990,
respectively. According the latter model during the ACRIM-gap  NIMBUS7/ERB  had to be shifted down by
0.58 $W/m^2$.

Finally,   Fr\"ohlich [2004, 2006] revised significantly his previous model correction
of  NIMBUS7/ERB data. He first acknowledged
that the supposed slip on May 1990 was indeed \emph{difficult to
really identify}. Then, he substituted the two step correction model
with a new model in which there was only one slip on 29 September
1989 followed by a upward linear trend. Figure  6  shows our analysis of these new
corrections in their latest version: on 29/09/1989 there is a
supposed slip of $+0.47 W/m^2$, which is significantly larger than
what was previously estimated, the upward trend is $+0.142Wm^{-2}/year$.
The total downward shift forced on NIMBUS7/ERB
record from 1989.5 to 1992.5 is about $0.86  W/m^2$, which is
significantly larger than what was previously estimated by Fr\"ohlich himself and the other groups.

\begin{figure}\label{NF-PMOD}
\includegraphics[angle=-90,width=20pc]{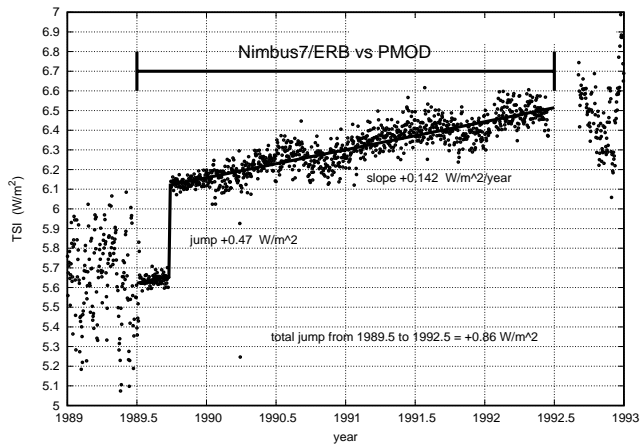}
 \caption{
Relative difference between  NIMBUS7/ERB original record and PMOD
TSI satellite composite.
 }
\end{figure}

From the above studies it is evident that  several opinions have
been formulated to solve the ACRIM-gap, even by the same research
team, and they quantitatively disagree with each other.
The above conflicting solutions indicate that it is not so certain  how
NIMBUS7/ERB should be corrected,   if some corrections are truly
needed.

Figure 7 shows our analysis of the comparison between
NIMBUS7/ERB and ERBS/ERBE. The 91-day moving average curve of
the relative difference between NIMBUS7/ERB and ERBS/ERBE  decreases
until August 1989 around the  time when SMM/ACRIM1 merges with
NIMBUS7/ERB at the level $6.29W/m^2$, as shown in the
graph. Since the beginning of September 1989 to the beginning of
1990 the curve rises rapidly by about $0.65W/m^2$. From 1990 to
1991.5 the curve rises by about $0.40W/m^2$. Finally, from 1991.5 to 1993 the curve decreases slightly by about $0.05W/m^2$.

\begin{figure}\label{NF-ERBS}
\includegraphics[angle=-90,width=20pc]{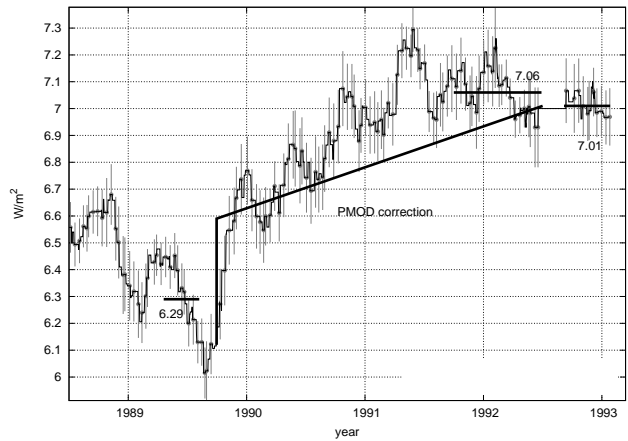}
 \caption{
91-day moving average of the relative difference between  NIMBUS7/ERB original record and ERBS/ERBE
TSI satellite composite during the ACRIM-gap. The figure shows the correction of NIMBUS7/ERB record implemented by PMOD team shown in Figure 6(solid line).
 }
\end{figure}

The total shift from 1989.5 to  1993 is about $0.72W/m^2$. Note that the error related to the single measurements is  about
$\pm0.11W/m^2$. Thus, the observed
difference between NIMBUS7/ERB and ERBS/ERBE is significant and must
be interpreted as due to biases in the data that are due to uncorrected
degradation problems in the sensors or something else that biases the TSI record.

Figure 7  shows also the correction implemented by the
PMOD team  on NIMBUS/ERB record [Fro\"ohlich 2004, 2006].
 It is evident that the PMOD team believes that the observed difference is due
 to uncorrected  problems occurring only on NIMBUS7/ERB's sensors. Even
 so, the correction of NIMBUS/ERB record implemented by the PMOD team ( $0.86  W/m^2$ from 1989.5 to 1992.5) appears to be  overestimated at least by about $0.09W/m^2$ because the total  shift observed during
  the period is no more than about $0.77W/m^2$. The difference seems to be due to the fact that
    PMOD team did not take into account that the real comparison must be done with the
     level  when SMM/ACRIM1 merges with NIMBUS7/ERB around the middle of 1989, and the
      level during this period, as indicated in the figure, is about $6.29W/m^2$. Thus, if
      on 29/09/1989 a jump really occurred in the NIMBUS7/ERB record, this has to be
      about $0.30W/m^2$, as previously estimated by Chapman et al. [1996] and Fro\"ohlich
       and Lean [1998]. Finally, the PMOD team's correction with a linear increase from
       29/09/1989 to 1992.5 is poorly observed in data shown in Figure 7; it appears to be just a linear simplification of the complex pattern observed in the figure.

The major problem with the interpretation of the
theoretical  studies [Lee III  \emph{et al.}, 1995; Chapman et al.,
1996; Fro\"ohlich and Lean, 1998; Fro\"ohlich, 2004, 2006] claiming that NIMBUS7/ERB is erroneous during the ACRIM-gap is that, although
these authors  did notice a difference between NIMBUS7/ERB and ERBS/ERBE
records, they  have interpreted such a difference as only due
to a corruption of the NIMBUS7/ERB record despite the fact that
ERBS/ERBE too was unable to continuously calibrate its sensor degradations and
its data had larger uncertainties than NIMBUS7/ERB data. Indeed, the increase
observed in  Figure 7  during the ACRIM gap could result from increased ERBS/ERBE
degradation relative to NIMBUS7/ERB, a relative increase in the
sensitivity of the NIMBUS7/ERB sensor, or both [Willson, 1997].

It is important to stress that in 1992 the experimental team responsible of  NIMBUS7/ERB record [Hoyt \emph{et al.}, 1992] corrected all biases in the data they could find and after that ever come up with a physical theory for the instrument that could cause it to become more sensitive. The NIMBUS7/ERB calibrations before and after the September 1989 shutdown gave no indication of any change in the sensitivity of the radiometer. When  Lee III \emph{et al.} of the ERBS team claimed there was an increase in NIMBUS7/ERB sensitivity, the NIMBUS7 team examined the issue and concluded there was no internal evidence in the NIMBUS7/ERB record to warrant the correction that the latter team was proposing (personal communication with Hoyt in Scafetta and Willson [2009]). Perhaps the increase between 1989 and 1991 in Figure 7 is indications of ERBS losing sensitivity rather than NIMBUS7 gaining sensitivity.

There are several  physical reasons to believe that
ERBS/ERBE could degrade  more likely than NIMBUS7/ERB in particular during the ACRIM-gap. For example: a)
The NIMBUS7/ERB cavity radiometer was in a relatively high altitude (about 900 km) while ERBS/ERBE was in a low earth orbit (ca. 200 km). It is possible that ERBS would degrade much faster than NIMBUS7/ERB due to more atmospheric bombardment of its sensor.
b) During the ACRIM-gap ERBS/ERBE was
experiencing for the first time the  enhanced solar UV radiation, which occurs
during solar maxima, and this too may have caused a  much faster degradation of the cavity coating of ERBS than of NIMBUS7/ERB because NIMBUS7 already experienced such degradation during the previous solar maximum; c) From the  spring 1990 to May/June 1991, when according to
Figure 7 the difference between NIMBUS7/ERB and
ERBS/ERBE increased by about $0.40W/m^2$, there was a rapid
 increase of cosmic ray flux, as Figure 8 shows. Also the latter phenomenon
might have more likely affected  ERBS/ERBE's sensors than
NIMBUS7/ERB's ones, which already experienced a solar maximum 10 years earlier.

Moreover, the cosmic ray count is negative-correlated to TSI and
magnetic flux, thus its minima correspond to solar activity maxima.
 Figure 8 shows that the minimum around 1991.5 was
lower than the minimum around 1989.8-1990.5. This implies that
according to this record  the solar activity was likely higher
around 1991.5 than around 1989.8-1990.5. This contradicts the
pattern observed in ERBS/ERBE while confirming NIMBUS7/ERB pattern,
as Figure 5 shows. However, other solar indexes, such as
the sunspot number index, present the opposite scenario. Thus,
it is unlikely that solar proxies indexes can be used to solve this issue definitely.

\begin{figure}\label{cosmic}
\includegraphics[angle=-90,width=20pc]{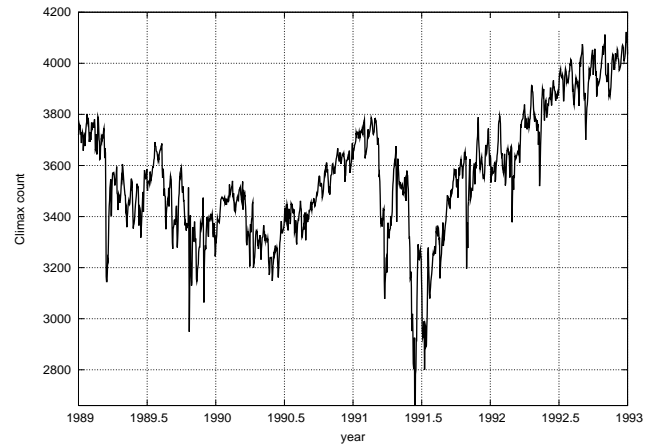}
 \caption{
Climax cosmic ray of  the University of New Hampshire (Version 4.8
21 December 2006). Data from
\emph{http://ulysses.sr.unh.edu/NeutronMonitor/}.
 }
\end{figure}

Thus, unless the experimental teams  find a physical theory for explaining the divergencies observed in their own  instrumental measurements and solve definitely the problem,     there exists  only a statistical way  to address  the ACRIM-gap problem by using the published data themselves. This requires just the acknowledgment of the existence of an unresolved uncertainty in the TSI satellite data. This can be done by:

1) Assuming that  NIMBUS7/ERB  is correct and  ERBS/ERBE is
erroneous; this would imply that during the ACRIM-gap  ERBS/ERBE record degraded  and should  be
shifted upward by  $0.72$-$0.77W/m^2$.

2) Assuming that   ERBS/ERBE is correct and  NIMBUS7/ERB is
erroneous; this would imply that during the the ACRIM-gap NIMBUS7/ERB
increased its sensitivity to TSI  and should  be
shifted downward by $0.72$-$0.77W/m^2$.

3) Assuming that both ERBS/ERBE and NIMBUS7/ERB records need
some corrections.

Note that there is no objective way to  implement method n. 3 and
infinitely different  solutions may be proposed. For example, the one proposed by the PMOD team is just one proposal among many others. Herein we
propose that all configurations between case n. 1 and case n. 2 may by possible, and for case n. 3 we just propose an average between the methods 1 and 2 stressing that
this arithmetic average should not be interpreted as a better
physical solution to the ACRIM-gap problem.

Figure  9  shows the three reconstructions of
NIMBUS7/ERB record in agreement with the above three scenarios: [A]
NIMBUS7/ERB data are unaltered; [C] the NIMBUS7/ERB data are altered
in such a way that their 91-day moving average curve in Figure
9  matches exactly the 91-day moving average curve of
ERBS/ERBE shown in Figure 5; finally, in [B] the
NIMBUS7/ERB data are altered in such a way that their 91-day moving
average  curve  matches exactly the average between the two  91-day
moving average curves of NIMBUS7/ERB and ERBS/ERBE shown in
Figure 5.

\begin{figure}\label{fig-3N7}
\includegraphics[angle=-90,width=20pc]{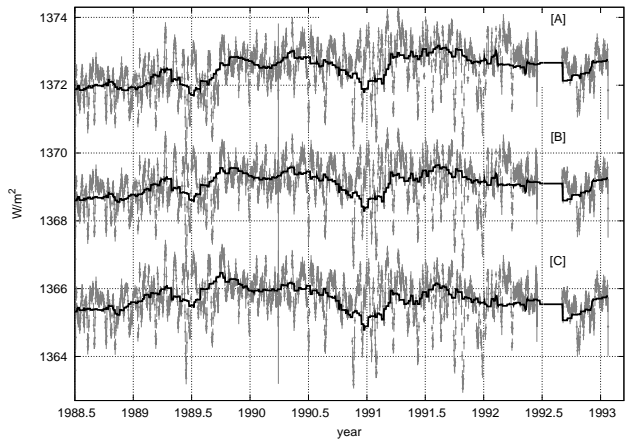}
 \caption{
Reconstructions of the NIMBUS7/ERB record during the ACRIM-gap in agreement with
 three alternative scenarios. [A] data are unaltered; [C] the data
 are adapted in such a way that their smooth component matches  exactly
 the smooth component of ERBS/ERBE; [B] the data
 are adapted in such a way that their smooth component matches
 exactly the average between the smooth components in [A] and  [C].
 }
\end{figure}

\section{SMM/ACRIM1 vs. UARS/ACRIM2}

To align SMM/ACRIM1 and UARS/ACRIM2 records we proceed as follows. First, we merge  NIMBUS7/ERB
record and its two alternative records  shown in Figure
9 with the SMM/ACRIM1 record. The merging is done by uniting the 91-moving average
mean curves at the merging day, 03/10/1991.

Second, we use the finding shown in Figure 10. This figure  shows the overlapping period between
NIMBUS7/ERB and UARS/ACRIM2 records. This interval is quite short and is made of
two separated intervals during which both satellite measurements
were interrupted for several months. Note that  the two intervals are not
aligned: there is a difference of about $0.2W/m^2$ between the two
levels. Because  the standard deviation of the data is about  $0.26W/m^2$
 which is significantly larger than the statistical error of measure
$0.16W/m^2$, the figure indicates that the data are not statistically
stationary.

\begin{figure}\label{fig-N7A2}
\includegraphics[angle=-90,width=20pc]{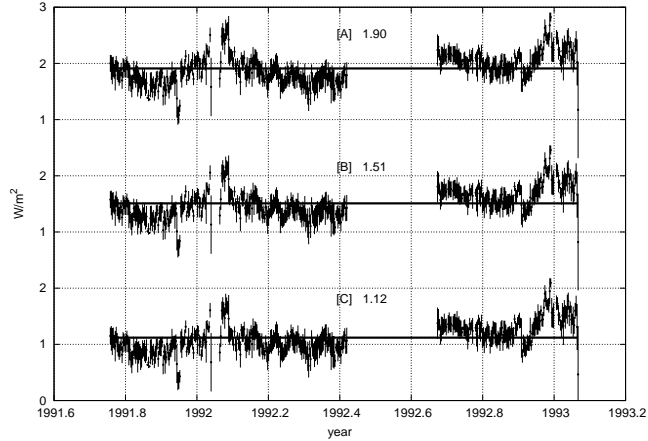}
 \caption{
 Relative difference between NIMBUS7/ERB and UARS/ACRIM2 TSI satellite records
 according to the three hypotheses discussed in the article. Note that NIMBUS7/ERB record has been already
 merged with SMM/ACRIM1 record. Thus, the value
 reported in the figure refers to the position SMM/ACRIM1 relative to UARS/ACRIM2 in the
  three alternative cases. }
\end{figure}

However, it is not evident which  is performing poorly: NIMBUS7/ERB and ERBS/ERBE or
UARS/ACRIM2. Because the difference observed between NIMBUS7/ERB and
UARS/ACRIM2 records in [A], and between the adapted NIMBUS7/ERB and
UARS/ACRIM2 records in [C] (where NIMBUS7/ERB record is adapted to reproduce the smooth of the ERBS/ERBE record) are almost equal, the first impression is that UARS/ACRIM2 sensors experienced a downward slip between the two intervals by about $0.2W/m^2$. However, because both NIMBUS7/ERB and ERBS/ERBE were less  able to calibrate their sensor degradation, it is still uncertain whether it is UARS/ACRIM2 record that has to be corrected and, if so, how large this corrections should be. Indeed, given the short time period and that both NIMBUS7/ERB and ERBS/ERBE are characterized by non stationary biases as large as $\pm0.2W/m^2$,  it is possible that during 1992 the two latter records experienced a similar upward bias. Thus, here we decided to keep UARS/ACRIM2 record unaltered and merge the two sequences using the average of the relative differences during the entire overlapping period in  all three cases,
as shown in the Figure 10. The error associated with
this merging is about $\pm0.1W/m^2$. However, if UARS/ACRIM2 record does need to be corrected, the global implication of this correction would be that the TSI satellite composite before 1992.5 should be  shifted downward by about $0.1 W/m^2$ in all three cases.

\section{UARS/ACRIM2 vs. ACRIMSAT/ACRIM3}

The merging between UARS/ACRIM2 and  ACRIMSAT/ACRIM3 is done  by
using the information shown in  Figure 11 that shows the
relative difference between ACRIMSAT/ACRIM3 and UARS/ACRIM2, and for
comparison, the relative difference between   SOHO/VIRGO and
UARS/ACRIM2. Note that the UARS/ACRIM2 measurements were interrupted
from 05/06/2001 to 08/16/2001.

\begin{figure}\label{fig-A32}
\includegraphics[angle=-90,width=20pc]{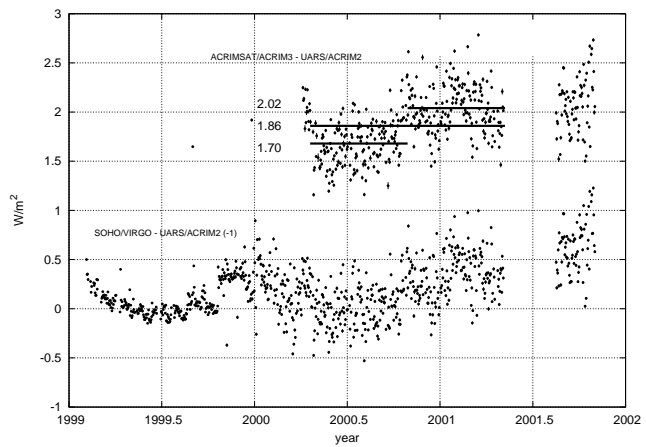}
 \caption{
 Relative difference between ACRIMSAT/ACRIM3 and UARS/ACRIM2 TSI satellite
 records (Top) and between SOHO/VIRGO and ACRIM2 TSI satellite records
 (Bottom). The latter is shifted down by $1W/m^2$ for visual clarity.
 }
\end{figure}

The latter comparison is necessary for determining the reason  of
the discrepancy observed between  ACRIMSAT/ACRIM3 and UARS/ACRIM2
which is significantly larger than the statistical error associated
with the measurements. In fact, the average statistical error of
UARS/ACRIM2 data is $0.01W/m^2$, while the average statistical error of
ACRIMSAT/ACRIM3 data is $0.008W/m^2$. The average statistical error
of the relative difference between ACRIMSAT/ACRIM3 and UARS/ACRIM2 is
no more than $0.018W/m^2$. However, the data in the  figure have a
standard deviation of about $0.3W/m^2$ which is significantly larger
than the statistical errors. Thus, the observed difference between
ACRIMSAT/ACRIM3 and UARS/ACRIM2 records is not due to random fluctuations, but to
non-stationary trends in the data.

Because a similar pattern appears when  UARS/ACRIM2 is compared
 with both SOHO/VIRGO and ACRIMSAT/ACRIM3 records, it is likely  that  UARS/ACRIM2
sensors  may have been experiencing some problem. Perhaps the annual cycle has been filtered off in some way. However, these  problems appear to have significantly modified   a natural
variation in the TSI data characterized by a time scale close to 1 year. Because the difference between
ACRIMSAT/ACRIM3 and UARS/ACRIM2 appears to present a cyclical pattern, an accurate way to merge the two
sequences is to evaluate the average during an entire period of
oscillation. The period from   04/05/2000 to 05/06/2001 covers
approximately one period of oscillation, and during this period the
average difference between ACRIMSAT/ACRIM3 and UARS/ACRIM2 is
$1.86W/m^2$: we use this value for the merging.  As the
figure shows the averages during the first and the second half of the
cycle are $1.70W/m^2$ and $2.02W/m^2$, respectively. This suggests that our merging
has an uncertainty of about $\pm0.18W/m^2$.

\section{Three updated ACRIM TSI composites}

The satellites records are merged and our three  TSI composites are
shown in Figure 12. Table 1 summarizes how SMM/ACRIM1 and UARS/ACRIM2 records have to be adjusted to be aligned with ACRIMSAT/ACRIM3.

\begin{figure}\label{fig.comp}
\includegraphics[angle=-90,width=20pc]{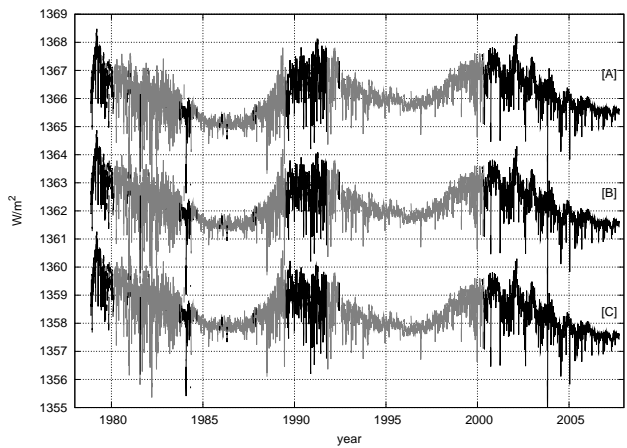}
 \caption{
The three TSI satellite conposites. The composites [B] and [C] are
shifted by $-4W/m^2$ and  $-8W/m^2$, respectively, for visual clarity.
 }
\end{figure}

The composite [A] shows that the 1996 minimum   is about $0.67\pm0.1
W/m^2$ higher than the 1986 minimum. The composite [B] shows that
the 1996 minimum   is about $0.28\pm0.1
W/m^2$ higher than the minimum in 1986. The composite [C] shows
that the 1996 minimum   is about $0.11\pm0.1 W/m^2$ lower than the minimum
in 1986.  Thus,  only in the eventuality that during the ACRIM-gap
ERBS/ERBE data are uncorrupted  the two solar
minima would  almost coincide, while on average the data indicate
that the TSI minimum in 1996 is  $0.28 \pm 0.4 W/m^2$ higher
than the minimum in 1985/6.

Note that if UARS/ACRIM2 record needs to be corrected during its superposition with NIMBUS7/ERB, as explained above, the TSI  1996 minimum relative to  the TSI 1986 minimum
would be  about $0.1 W/m^2$ higher  than the above three estimates. Thus, if this is the case, according to the satellites data the difference between the two minima would be  about $0.38 \pm 0.4 W/m^2$. This would further stress that the TSI satellite data do indicate that TSI likely increased during solar cycles 21-23 (1980-2002).

\begin{center}
\begin{table}
\begin{tabular}{|c|c|c|c|}
  \hline
   & [A] & [B] & [C] \\\hline
  A1 & $-1.90\pm0.19$ & $-1.51\pm0.19$ & $-1.12\pm0.19$ \\\hline
  A2 & $+1.86\pm0.16$ & $+1.86\pm0.16$ & $+1.86\pm0.16$ \\\hline
  A3 & $0$ & $0$ & $0$ \\
  \hline
\end{tabular}
\caption{
 Position in $W/m^2$ of SMM/ACRIM1 (A1) and UARS/ACRIM2 (A2) relative to ACRIMSAT/ACRIM3 (A3) in the three scenarios [A], [B] and [C] as discussed in the test.
 The errors are calculated by taking into account the highest  uncertainty due to the statistical non-stationarity of NIMBUS7/ERB and UARS/ACRIM2 when they merge and when UARS/ACRIM2 merges ACRIMSAT/ACRIM3. Note that the error of the position of SMM/ACRIM1 compared to UARS/ACRIM2 is $\pm0.1W/m^2$. The error associated to the non-stationarity of NIMBUS7/ERB during the ACRIM-gap is described by the three scenarios [A], [B] and [C].
 }
\end{table}
\end{center}

\section{TSI proxy secular reconstructions}

  It is necessary to use reconstructions of the solar activity as long as possible, at least one century, for determining the effect of solar variations on climate. The TSI record that is possible to obtain from direct TSI satellite measurements covers the period since 1978, and this period is far too short to correctly estimate how the Sun may have altered climate. The reason is because the climate system is characterized by a slow characteristic time response to external forcing that is estimated to be about 8 years (which,  theoretically, can be as large as 12 years) [Scafetta, 2008; Schwartz, 2008]. This decadal time response of the climate requires several decade long records for a correct evaluation of an external forcing on climate.
Thus, it is necessary to merge the TSI satellite composites with the long TSI secular reconstructions, which are quite uncertain because they are necessarily based on proxy data, and not direct TSI measurements.

Long-term TSI changes over the past 400 years since the 17th-century Maunder minimum
have been reconstructed by several authors, for example:  Hoyt and Schatten [1997],
 Lean [2000], Wang \emph{et al.} [2005] and  Krivova \emph{et al.} [2007]. These TSI proxy reconstructions
 are  based on the sunspot number record, the long-term trend in geomagnetic
activity,  the solar modulation of cosmogenic isotopes such as $^{14}C$ and $^{10}Be$ records, and other solar related records. These observables are used because they are supposed to be linked to TSI variations. However, it is not known exactly how the TSI can be reconstructed  from these historical records nor whether these records are sufficient to faithfully reconstruct TSI changes. Thus, the proposed TSI secular proxy reconstructions are quite different from each other and show different patterns, trends and maxima, as depicted in Figure 13. Nevertheless, they  reproduce similar patterns: in particular, note the minima during the Maunder Minimum (1645-1715) and the Dalton Minimum (1790-1820), and the TSI increase during the first half of the 20$^{th}$ century.

\begin{figure}\label{fig.comp}
\includegraphics[angle=-90,width=20pc]{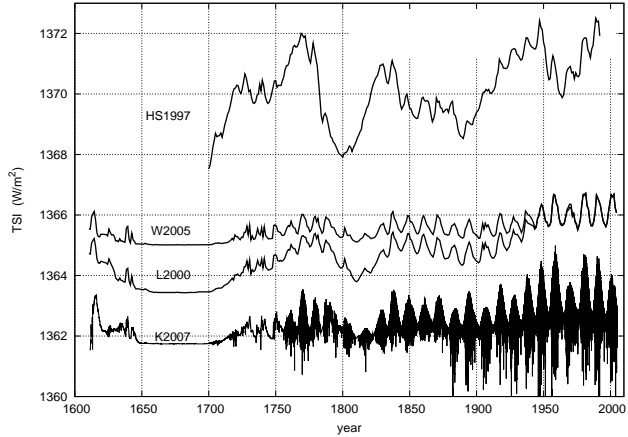}
 \caption{
Secular  TSI proxy reconstructions by Hoyt and Schatten [1997] (HS1997),
 Lean [2000] (L2000), Wang \emph{et al.} [2005] (W2005) and  Krivova \emph{et al.} [2007] (K2007). K2007 has been shifted by $-3W/m^2$ for visual convenience.
 }
\end{figure}

The TSI increase during the first half of the 20$^{th}$ century is particularly important. In fact,  because the characteristic time response of climate to external forcing is  about 8-12 years, an increase of TSI during the first half of the  20$^{th}$ century would induce a warming also during the second half of the  20$^{th}$ century, even if the TSI  remains almost constant during the second half  of the  20$^{th}$ century [Scafetta and  West, 2007].

The four TSI proxy reconstructions shown in Figure 13 present different trends since 1975. The TSI  reconstruction by Hoyt and Schatten [1997] suggests that TSI increased during this period, as shown in our TSI satellite composites [A] and [B], and in the original ACRIM TSI satellite composite. However, the other three TSI proxy reconstructions [Lean, 2000; Wang \emph{et al.}, 2005;  Krivova \emph{et al.}, 2007] suggest that TSI did not change on average since 1978, as shown in our  TSI satellite composite [C] and in the PMOD TSI satellite composite. Thus, the uncertainty that we have found in composing the TSI satellite records appears unresolved also by using the TSI proxy reconstructions  because different solar proxies do suggest different TSI patterns as well.

Because the TSI satellite composites refer to the actual TSI measurements, we propose their merging with the TSI proxy reconstructions for obtaining a TSI secular record. Here, we chose the most recent TSI proxy reconstruction [Krivova \emph{et al.}, 2007], which has  a daily resolution, and merge it to the TSI satellite composites in such a way that their 1980-1990 average coincides.
Other choices and their implications by using the original ACRIM and the PMOD TSI satellite composites with the TSI proxy reconstructions of Lean  [2000] and  Wang \emph{et al.} [2005] can be found in Scafetta and West [2007]. The three  TSI merged records herein proposed are shown in Figure 14. The figure shows that during the last decades the TSI has been at its highest values since the 17$^{th}$ century.

\begin{figure}\label{fig.comp}
\includegraphics[angle=-90,width=20pc]{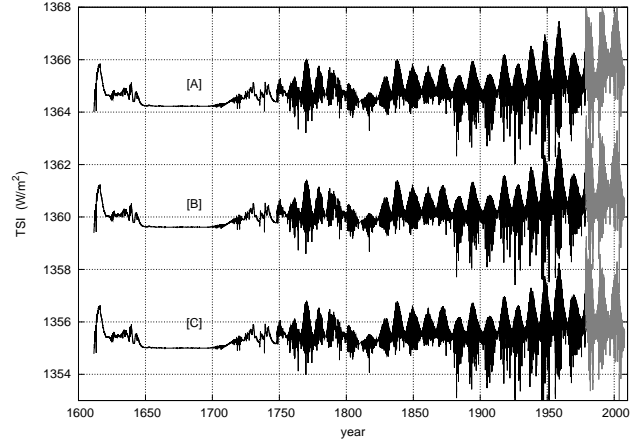}
 \caption{
Merging of the secular  TSI proxy reconstruction by Krivova \emph{et al.} [2007] (black) with the three TSI satellite composites proposed in Figure 12 (grey). The TSI reconstructions [B] and [C] are shifted by $-5W/m^2$ and $-10W/m^2$, respectively, for visual convenience. The merging is made by shifting the original secular  TSI proxy reconstruction by Krivova \emph{et al.} [2007] by $-0.5174W/m^2$, $-0.1295W/m^2$ and $+0.2584W/m^2$ in the case [A], [B] and [C] respectively.
 }
\end{figure}

\section{Phenomenological solar signature on climate}

The phenomenological solar signature on climate can be estimated with a
phenomenological energy balance model (PEBM) [Scafetta and West, 2007].
PEBM  assumes that the climate system, to the lowest-order
approximation, responds to an external radiative forcing as a simple
thermodynamical system, which is characterized by a given relaxation
time response $\tau$ with a sensitivity $\alpha$. The physical meaning of it is
that a small anomaly (with respect to the TSI average value) of the
solar input, measured by $\Delta I$, forces the climate to  reach a
new thermodynamic equilibrium at the asymptotic temperature value
$\alpha \Delta I$ (with respect to a given temperature average
value). Thus, if $\Delta I(t)$ is a small variation (with respect to
a fixed average) of an external forcing and $\Delta T_s(t)$ is the
Earth's average temperature anomaly induced by $\Delta I(t)$,
$\Delta T_s(t)$ evolves in time as:
\begin{equation}\label{prrm}
    \frac{d\Delta T_s(t)}{dt}=\frac{\alpha \Delta I(t)-\Delta
    T_s(t)}{\tau}.
\end{equation}
 A model equivalent to (\ref{prrm}) has
been used as a basic energy balance model [\emph{North et al.},
1981; \emph{Douglass and Knox}, 2005], but herein we use TSI records
as a \emph{proxy} forcing.

We implement the  PEBM by imposing  that the global peak-to-trough amplitude
of the  11-year solar cycle signature on the surface temperature is about $0.1K$
 from 1980 to 2002, as found by several authors (see IPCC [2007], page 674 for details).
 This implies that the climate sensitivity $Z_{11}$ to the 11-year solar cycle is
\begin{equation}\label{cs}
   Z_{11}=0.11\pm0.02 K/Wm^{-2},
\end{equation}
 as found by Douglass
and Clader [2002], and Scafetta and West [2005]. In addition the characteristic time response to external forcing  has been phenomenologically estimated to be $\tau=8\pm2$ years [Scafetta, 2008; Schwartz, 2008]. Note that Scafetta [2008] has also found that climate is characterized by two characteristic time constants $\tau_1=0.40\pm0.1$ and $\tau_2=8\pm2$, with the latter estimate that may be a lower limit (the upper limit being  $\tau_2=12\pm3$ years), but a discussion about the consequences of this finding is left to another study [Scafetta, 2009].

The value of the parameter $\alpha$ is not calculated theoretically by using the TSI as a climate forcing as usually done in the traditional climate models. The value of $\alpha$ is calculated  by using the phenomenological climate sensitivity to the 11-year solar cycle found in Eq. [\ref{cs}] by means of the following equation
 \begin{equation}\label{tc}
  \alpha(\tau)=Z_{11}\sqrt{1+\left(\frac{2\pi\tau}{11}\right)^2},
\end{equation}
which solves Eq. [\ref{prrm}]. Thus, we find that the phenomenological climate sensitivity to TSI changes is
\begin{equation}\label{aa}
   \alpha=0.51 \pm  0.15 K/Wm^{-2}.
\end{equation}
With the above value of $\tau$ and $\alpha$ Eq. [\ref{prrm}] can be numerically solved by using as input the TSI records shown in Figure 14. The phenomenological solar signatures (PSSs) are shown in Figure 15 where the three PSSs are plotted since 1600 against a paleoclimate Northern Hemisphere temperature reconstruction [Moberg \textit{et al.}, 2005] and since 1850 against the actual instrumental Northern Hemisphere surface record [Brohan  \emph{et al.}, 2006].

The figure shows that there is a good agreement between the PSSs  and the temperature record. The patterns between 1600 and 1900 are well recovered. The warming during the first half of the 20$^{th}$ century is partially recovered. Finally, since 1978 the output strongly depends on the TSI behavior. If the TSI reconstruction [A] is adopted, a significant portion of the warming, about 66\% observed since 70s has been induced by solar variations, while if the TSI reconstruction [C] is adopted, almost all warming, about 85\% observed since 70s has been induced by factors alternative to solar variations. If the average TSI reconstruction [B] is adopted, at most 50\% of the warming observed since 70s has been induced by solar variations.

\begin{figure}\label{fig.comp}
\includegraphics[angle=-90,width=20pc]{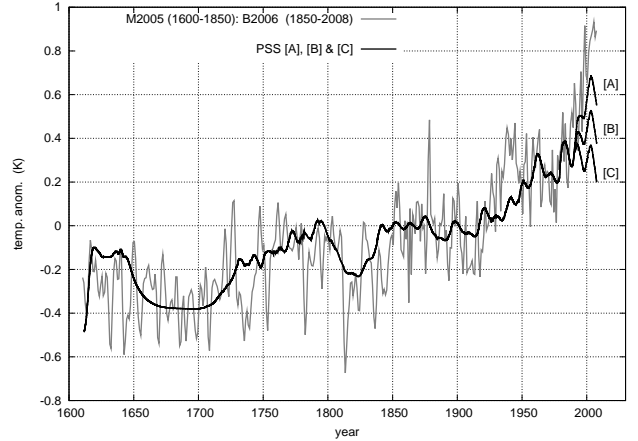}
 \caption{
The three phenomenological solar signatures on climate (black) obtained with Eq. \ref{prrm} forced with the three TSI records shown in Figure 14 against a paleoclimate Northern Hemisphere temperature reconstruction [Moberg \textit{et al.}, 2005] from  1600 to 1850, and since 1850 against the actual instrumental Northern Hemisphere surface record [Brohan  \emph{et al.}, 2006] (grey). The figure shows temperature anomalies relative to the 1895-1905 average.
 }
\end{figure}

\section{Conclusion}

We have reconstructed  new TSI satellite composites by using the
three ACRIM records. We have shown that different composites
are possible depending on how the ACRIM-gap  from 1989.5 to
1992 is solved.
Our three TSI composites indicate that the TSI minimum in 1996 is at least
approximately $0.30 \pm 0.40 W/m^2$ higher than the TSI minimum in
1986. And that the two minima would approximately be located at the same level only
in the eventuality that the TSI ERBS/ERBE satellite record is uncorrupted during the ACRIM-gap, a fact that may be not likely.

None of the TSI satellite composites proposed by the  ACRIM, IRMB and PMOD teams can be considered rigorously correct. All three teams have just adopted alternative methodologies that yield to different TSI composites, but these teams have ignored  the unresolved uncertainty in the data that yields to an unresolved uncertainty in the TSI composites as well.

Note that comparison with theoretical TSI proxy models, for example Wenzler \emph{et al.}, [2006] and Krivova \emph{et al.} [2007], cannot be used to resolve the issue, as the PMOD team assumes, because: 1) In science theoretical models have to be tested against the observations, not vice versa; 2) The TSI proxy models adopt a reductionistic scientific approach, that is, they assume that some given solar observable that refers to a \emph{particular} solar measure (for example measurements from magnetograms or measurements of the intensity of a given frequency of the spectrum) can be used to faithfully reconstruct a \emph{global} solar measure such as the TSI; 3) The TSI proxy models do depend on parameters that opportunely calibrated  give different outcomes that can, eventually, fit alternative satellite composites.

Thus, because it is not possible to reconstruct with certainty  the TSI behavior during the ACRIM-gap, the TSI decadal trend during the last three decades is unfortunately uncertain, and any discussion that needs to use the TSI record has to take into account this  unresolved uncertainty.

However, because the uncertainty in the data  indicate that the TSI minimum in 1996 is at least
approximately $0.30 \pm 0.40 W/m^2$ higher than the TSI minimum in
1986,
on average   the satellite records do suggest that TSI may have increased
from 1980 to 2000. Therefore, the sun may have  significantly contributed
to the warming observed during the last three decades, as suggested by the phenomenological energy balance model simulations herein proposed.

Note that a recent paper by Lockwood [2008]
concludes that even with the adoption of the original ACRIM composite, the
sun's contribution to the global surface warming would be negligible
during the last three decades, in contrast with the findings of
Scafetta and West [2007,2008] and those presented here. However,
Lockwood's findings derive from his evaluation of the characteristic time response of the climate
 to solar variation: $\tau=0.8$ years. This value strongly differs from the value herein adopted of $\tau=8$ years and recently measured by Scafetta [2008] and  Schwartz [2008].
 The problem with Lockwood's short time constant  is that according to the climate physics implemented
 in most climate models,
 the characteristic time response of the climate
 varies from a few months  to several
years and even decades, as Lockwood himself acknowledges in his paper (see references there). For example, the linear upwelling/diffusion energy balance model used by Crowley [2000] is characterized by a time response of about $\tau=10$ years. 
In addition, Scafetta [2008] and Schwartz  [2008] have found that climate is indeed characterized by two characteristic time constants, one short with a time scale of several months and one long with a decadal time scale.
 The  climate processes with a fast response
are usually responsible for the fast fluctuations seen in the data. Instead,  the climate processes with a slow response are those that drive the
decadal and secular trends observed in the global temperature. This
slow climate response derives from the fact that the processes that
regulate the decadal and secular variation of climate (most of all
energy exchange with the deep ocean  and changes of the albedo due
to the melting of the glaciers and forestation and desertification
processes) are very slow processes, and they work as powerful climate
feedbacks. Thus, we believe that Lockwood's analysis is inappropriate because it failed to take into account  the climate processes with a slow time response that would be responsible of a strong climate response to solar changes. However, a more detailed discussion about this issue, which would imply also an update of the PEBM presented here, is left to another study [Scafetta, 2009].

\textbf{Acknowledgment:} NS thanks the Army Research Office for research
support (grant W911NF-06-1-0323).

\end{document}